\documentclass[aps,prl,showpacs,twocolumn,floatfix]{revtex4}
\usepackage{graphicx,graphics,epsf,times,amsmath,amsfonts,amssymb,latexsym,bm,bbm}

\newcommand{\ignore}[1]{}

\begin{document}

\title{Quantum Adiabatic Brachistochrone}
\author{A. T. Rezakhani$^{(1)}$, W. -J. Kuo$^{(1)}$, A. Hamma$^{(2)}$, D. A. Lidar%
$^{(1)}$, and P. Zanardi$^{(1)}$}
\affiliation{$^{(1)}$Departments of Chemistry, Electrical Engineering, and Physics, and
Center for Quantum Information Science \& Technology, University of Southern
California, Los Angeles, CA 90089, USA}
\affiliation{$^{(2)}$Perimeter Institute for Theoretical Physics, 31 Caroline St. N, N2L
2Y5, Waterloo ON, Canada}

\begin{abstract}
We formulate a time-optimal approach to adiabatic quantum computation (AQC).
A corresponding natural Riemannian metric is also derived, through which AQC
can be understood as the problem of finding a geodesic on the manifold of
control parameters. This geometrization of AQC is demonstrated through two
examples, where we show that it leads to improved performance of AQC,
and sheds light on the roles of entanglement and curvature of the
control manifold in algorithmic performance.
\end{abstract}

\maketitle

\textit{Introduction.---}Quantum computation is most commonly formulated in
the language of the \textquotedblleft circuit model\textquotedblright\ \cite%
{Nielsen-book}. The problem of finding optimal quantum circuits -- which
minimize the number of gates used -- was recently addressed in an elegant
differential-geometric framework, wherein the minimum number of elementary
gates for construction of a general $n$-qubit unitary $U$ can be found by
traversing the geodesic connecting the identity $\openone$ to $U$ over the $%
\text{SU}(2^{n})$ manifold \cite{Nielsen-science}. This approach is
appealing since it allows for the application of powerful tools and
techniques from
variational calculus and
differential geometry \cite{Nakahara-book} to quantum computation. Adiabatic quantum computation (AQC) 
\cite{Farhi}, on the other hand, is a very different approach which has
recently attracted much attention due to its fundamental connection to
quantum many-body systems, in particular to
quantum phase transitions (QPTs) \cite{AQCmanybody}. The basic strategy of AQC is to solve
computational problems based on
adiabatic evolution. A quantum system is
prepared in the ground state of an initial Hamiltonian $H(0)=H_{I}$. The
system is then driven to a final state, which -- provided the evolution was
adiabatic -- is the ground state of a \textquotedblleft problem
Hamiltonian\textquotedblright\ $H(T)=H_{P}$. This final state 
corresponds to the solution of a hard problem, while a short time $T$
corresponds to an efficient AQC strategy. Although AQC is equivalent in
computational power to the circuit model \cite{AQCeq}, and relies on one of
the oldest theorems of quantum theory -- the adiabatic theorem \cite%
{Messiah-book} -- it is still relatively unexplored. Specifically, an
optimal strategy for AQC, akin to what has been done for the circuit model 
\cite{Nielsen-science}, has not yet been formulated.

In this work we reformulate AQC as a variational problem and develop a
time-optimal strategy -- a \textquotedblleft quantum adiabatic
brachistochrone\textquotedblright\ (QAB) -- for quantum algorithms \cite%
{Carlini-timeopt}. Specifically, we devise a variational time-optimal
strategy for obtaining an interpolating Hamiltonian $H(t)$ between $H_{I}$
and $H_{P}$, which gives rise to the shortest time $T$ while guaranteeing
that the actual final state (the solution to the corresponding Schr\"{o}%
dinger equation), is close to the desired final ground state.
We go further and show that the QAB can be recast in a natural
differential-geometric framework. Specifically, we construct a Riemannian
geometry, along with its corresponding metric, for the adiabatic evolution.
We provide two examples which illustrate the advantage of this optimal
approach.

\textit{Time-optimal AQC.---} The adiabatic approximation, which underlies
AQC, is often stated as follows \cite{Messiah-book}. Consider a system
subjected to a time-dependent Hamiltonian $H(t)$, with a nondegenerate
ground state $|\Phi _{0}(t)\rangle $ isolated by a nonvanishing gap $\Delta
(t)$ from the first excited state $|\Phi _{1}(t)\rangle $. Let $D(t)\equiv
|\langle \Phi _{1}(t)|\partial _{t}H(t)|\Phi _{0}(t)\rangle |$. Prepare the
system in $|\psi (0)\rangle =|\Phi _{0}(0)\rangle $ and let it evolve
according to the Schr\"{o}dinger equation into the state $|\psi (T)\rangle $%
. Then, provided the time variation of the Hamiltonian is sufficiently slow,
or $T$ is sufficiently large, in that $\max_{t\in \lbrack
0,T]}D(t)/\min_{t\in \lbrack 0,T]}\Delta ^{2}(t)\ll \epsilon $, the fidelity 
$F(T)\equiv |\langle \Phi _{0}(T)|\psi (T)\rangle |$ between the final
ground state and the actual final state is high: $F\geq \sqrt{1-\epsilon ^{2}%
}$. However, it is well known that the latter condition is not always
accurate \cite{ATcrit}, as recently verified experimentally \cite{Suter}.
Rigorous versions of the adiabatic approximation \cite{ATrigorous} typically
involve $\Vert \partial _{t}H\Vert ^{2}/\Delta ^{3}$ (with the norm being
the maximum eigenvalue), or terms with different powers of $\Vert \partial
_{t}H\Vert $ and $\Delta $. 
As our approach is to use the adiabatic condition as a heuristic for finding
optimal trajectories, the exact form of the adiabatic condition is in fact
not essential: we shall judge success by the tradeoff between fidelity $F$
and evolution time $T$. As we show below, this pragmatic approach also
allows us to find time-optimal and geometric formulations of AQC.

The time-dependence of Hamiltonians usually comes from a set of control
parameters $\mathbf{x}(t)=\left( x^{1}(t),\ldots ,x^{M}(t)\right) $ -- e.g.,
electric or magnetic fields, laser beams, or any other experimental
\textquotedblleft knob\textquotedblright\ -- varying over a parameter
manifold $\mathcal{M}$, whence $H=H[\mathbf{x}(t)]$. 
Varying the Hamiltonian for a given interval $t\in \lbrack t_{0},t_{1}]$
then translates geometrically into moving along a control curve (or path) $%
\mathbf{x}(t)$ in $\mathcal{M}$. We can reparameterize $\mathcal{M}$ via a
dimensionless \textquotedblleft natural parameter\textquotedblright\ $s(t)$ 
\cite{Nakahara-book}, with $s(0)=0$ and $s(T)=1$ (e.g., the normalized
length), where $v(t)\equiv \mathrm{d}s/\mathrm{d}t>0$
characterizes the speed by which we move along $\mathbf{x}[s(t)]\in \mathcal{%
M}$. To make the adiabatic dynamics \textit{locally} compatible with the
geometric structure of $\mathcal{M}$, we modify 
the adiabatic condition into the following local (i.e., instantaneous) form 
\cite{RolandCerf}: 
\begin{equation}
\frac{v(s)\Vert \partial _{s}H(s)\Vert }{\Delta ^{2}(s)}\ll \epsilon \quad
\forall s\in \lbrack 0,1].  \label{eq:T-local}
\end{equation}%
Hereafter the norm is the Hilbert-Schmidt (or Frobenius) norm, defined as 
$\Vert A\Vert _{\text{HS}}=\sqrt{\text{Tr}[A^{\dag }A]}$. From the 
relation $T=\int_{0}^{1}\mathrm{d}s/v(s)$, 
we define the \emph{adiabatic time-functional} 
\begin{equation}
\mathcal{T}[\mathbf{x}(s)]=\int_{0}^{1}\frac{\mathrm{d}s}{v_{\text{ad}}[\dot{%
\mathbf{x}}(s),\mathbf{x}(s)]}\equiv \int_{0}^{1}\mathrm{d}s~\mathcal{L}[%
\dot{\mathbf{x}}(s),\mathbf{x}(s)],
\end{equation}%
where $\dot{x}\equiv \partial _{s}x$.
Inspired by the local adiabatic condition (\ref{eq:T-local}), we
choose the instantaneous ``adiabatic speed'' via the \textit{ansatz} 
\begin{equation}
v_{\text{ad}}(s)\equiv \epsilon \Delta ^{2}(s)/\Vert \partial _{s}H(s)\Vert ,
\label{eq:v-ad}
\end{equation}%
and hence -- using Einstein summation -- the Lagrangian
$\mathcal{L}[\dot{\mathbf{x}}(s),\mathbf{x}(s)]=\Vert
\dot{x}^{i}\partial _{i}H[\mathbf{x}(s)]\Vert /\epsilon \Delta
^{2}[\mathbf{x}(s)]$, where 
$\partial_i\equiv\partial/\partial x^i$. This ansatz is
sensible, in that adiabaticity is hindered when the gap closes, while it is
favored when the variation of the Hamiltonian is slow. 
To simplify the analysis, from now on, we take the Lagrangian to be
$\mathcal{L}'=\mathcal{L}^2$;
this corresponds to a reparameterization
which leaves the length of the curve solving the Euler-Lagrange (EL) equations
invariant \cite{Nielsen-science,Nakahara-book}.

$\mathcal{T}$ measures the time taken to traverse the curve $\mathbf{x}(s)$
from start to finish, subject to the local adiabatic condition (\ref%
{eq:T-local}). Our goal is to minimize this time and thus obtain the \emph{%
time-optimal curve}, or set of time-dependent controls. This optimal curve
is the QAB. From variational calculus, the optimal path $\mathbf{x}_{\text{%
QAB}}(s)$ should satisfy $\delta \mathcal{T}[\mathbf{x}(s)]/\delta \mathbf{x}%
(s)=0$, which gives rise to the
EL equations $\mathrm{d}\partial
_{\dot{\mathbf{x}}}\mathcal{L}'[\dot{\mathbf{x}},\mathbf{x}]/\mathrm{d}%
s=\partial _{\mathbf{x}}\mathcal{L}'[\dot{\mathbf{x}},\mathbf{x}]$. 

\begin{figure*}[tp]
\includegraphics[width=5.3cm,height=3.2cm]{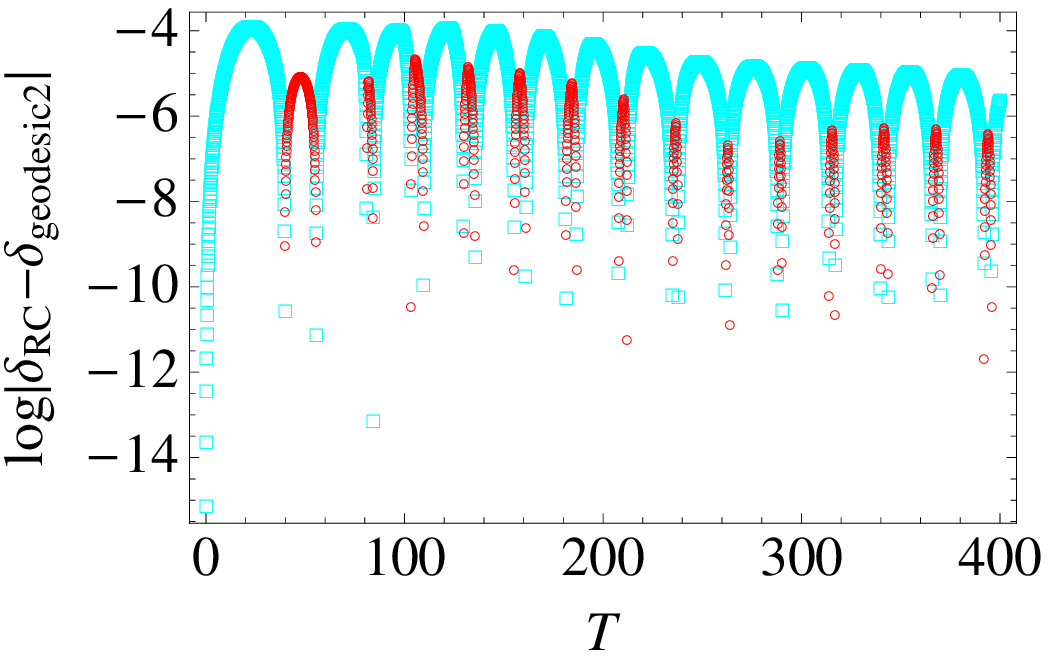} \hskip 3mm
\includegraphics[width=5.3cm,height=3.2cm]{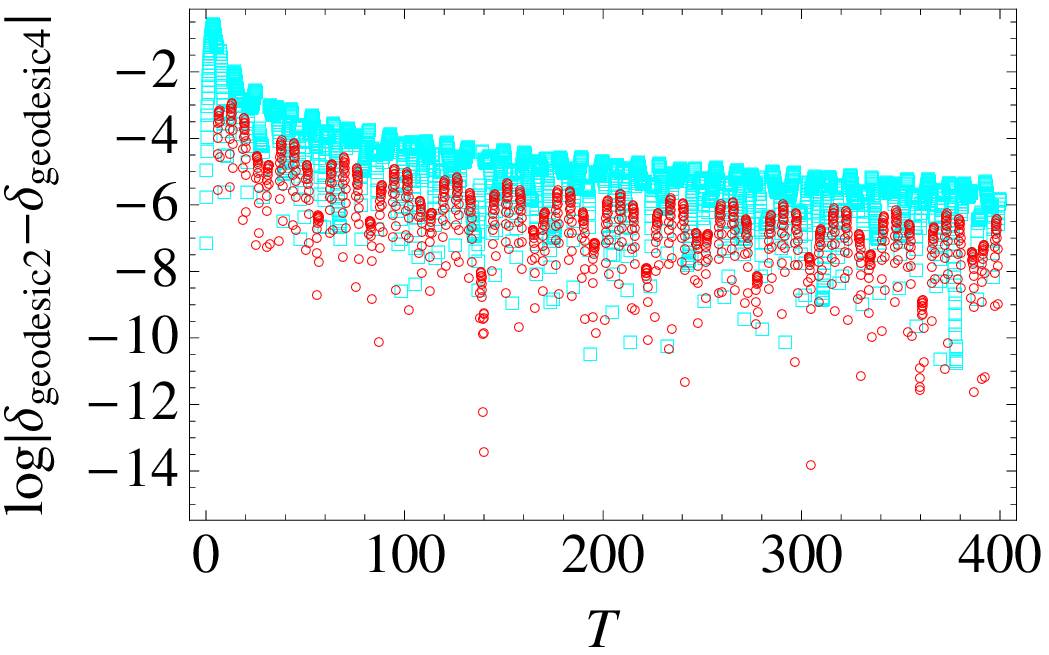} \hskip 3mm
\includegraphics[width=4.9cm,height=3.2cm]{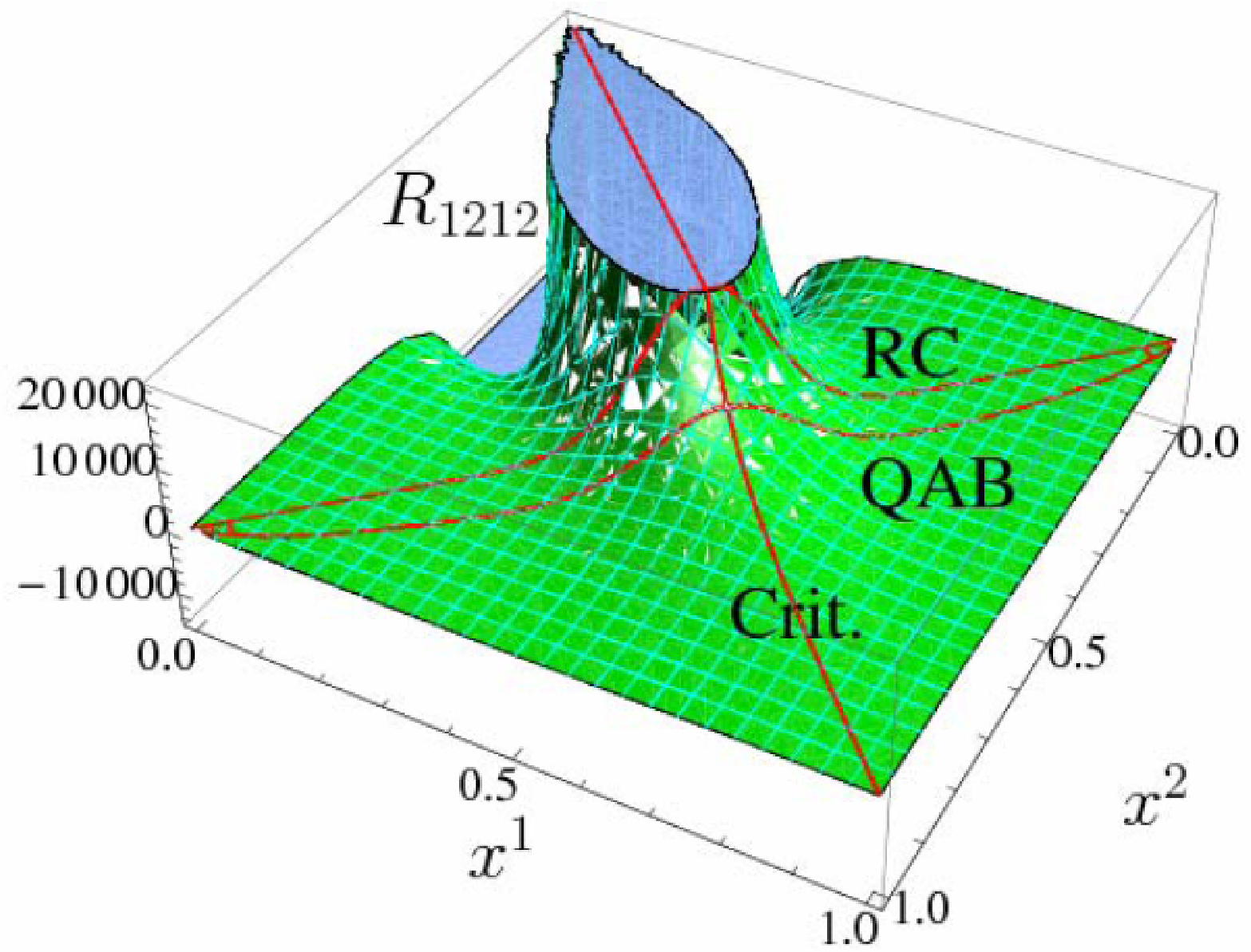} 
\caption{(color online) Left: Final-time error $\protect\delta(T)$ for the
RC and 2-d geodesic paths for the Grover search problem, for $n=6$ qubits.
Squares (cyan) indicate where the 2-d geodesic path outerperforms the RC
path ($\protect\delta_{\text{geodesic}}\leq\protect\delta_{\text{RC}}$);
circles (red) correspond to the opposite case. Oscillations are due to $%
\protect\delta(T)$ itself being highly oscillatory, with an envelope $%
\propto 1/T$. Middle: $\protect\delta(T)$ for the 2-d and 4-d geodesic
paths, for $n=1$. 
Cyan squares (red circles) indicate where the 4-d (2-d)
geodesic path results in a smaller error. Right: Component
$R_{1212}(x^1,x^2)$ of the curvature tensor for $n=3$. The curves on
the curvature surface 
show the critical line (vanishing gap as $n \rightarrow \infty$), the
RC interpolation, and the 2-d geodesic (QAB).
$R_{1212}$ is the only independent component of the curvature tensor
in this case
[$R_{ijkl}=2R_{1212}(g_{ik}g_{jl}-g_{il}g_{jk})/\det\mathbf{g}$].} 
\label{comparison}
\end{figure*}

Some remarks regarding the QAB are in order. (i) The total real evolution
time $T$ (for a given fidelity $F$) is not necessarily the same as the
time-functional $\mathcal{T}$. The correct interpretation is this: after
finding the optimal path $\mathbf{x}_{\text{QAB}}(s)$ we
solve the Schr\"{o}dinger equation 
$i v_{\text{ad}}\partial _{s}|\psi \rangle =H|\psi \rangle $
with $H$ and $v_{\text{ad}}$ computed along the optimal path, 
to find $|\psi \lbrack \mathbf{x}_{\text{QAB}}(t)]\rangle $.
The actual adiabatic error is then $\delta (T)\equiv \sqrt{1-F_{\text{QAB}%
}(T)^{2}}$, where $F_{\text{QAB}}(T)\equiv |\langle \Phi _{0}(T)|\psi
\lbrack \mathbf{x}_{\text{QAB}}(T)]\rangle |$, for a given $T$. With this
interpretation it is safe to take $\epsilon \equiv 1$ in Eq.~(\ref{eq:v-ad}%
). Thus, we use $\mathcal{T}$ as a guiding principle to obtain optimal
paths; the actual adiabatic time and error should be calculated
independently, as per the prescription above. (ii) Equation~(\ref{eq:T-local}%
) and our corresponding choice of $v_{\text{ad}}$ are by no means unique.
They merely represent a convenient ansatz for our subsequent analysis, and
it is quite possible that a better ansatz involving a different combination
of $\Vert \partial _{s}H\Vert $ and $\Delta $ exists. (iii) The presence of $%
\Delta $ in $\mathcal{L}$ implies that in order to apply our method, one
needs to either find the gap exactly (which could be as hard as solving the
problem itself), or restrict the interpolation to forms for which an
explicit functional form for the gap can be obtained (e.g., in exactly- or
almost-exactly-solvable models), or follow the scheme suggested to translate
a quantum circuit model computation to an AQC (by which an interpolating
Hamiltonian with an easily calculable gap is derived) \cite{AQCeq}, or
estimate the gap by means of other methods (e.g.,
experimentally).
(iv) The bottleneck of adiabatic algorithms is at the finite-size
precursor of QPTs where the gap becomes small (and vanishes in the
thermodynamic limit) \cite{AQCmanybody}.
The subtle point, however, is that due to the $\Vert \dot{x}%
^{i}\partial _{i}H\Vert $ factor in the numerator of $\mathcal{L}$, in
principle, there is the possibility that one can (at least partially)
suppress the vanishing gap and associated QPT \cite{Avoid}. The QAB will
inherently seek to identify such criticality-suppressing control strategies
(relative to the specific adiabatic condition we have adopted here), if they
exist.
Indeed, we shall see this in our examples, below.
(v) In general, not every choice for the norm $\Vert \cdot \Vert $
yields an analytic adiabatic velocity. Our choice of the Hilbert-Schmidt
norm is made to ensure analyticity, and to simplify our calculations. We
shall see below that this choice also enables a geometric treatment of the
QAB problem. (vi) 
Modeling AQC usually necessitates a parameterization of the
Hamiltonian. In general, though, this parametrization is not unique. 
For simplicity, 
consider the parametrization $H(\mathbf{x})=\sum_{i}x^{i}\sigma _{i}$,
where $\{\sigma _{i}\}_{i=1}^{M}$ are 
time-independent, noncommuting, linearly-independent Hermitian operators,
chosen in accordance with the underlying structure of the optimization or
physical problem in question. For example, in a multi-qubit system, $\{\sigma _{i}\}$ could
represent (two-) local interactions in the form of the tensor
product of Pauli matrices.
It can be shown that in these ``interaction coordinates,'' the
EL equations read: 
\begin{equation}
\ddot{x}^{k}+\Gamma _{~~ij}^{k}(\mathbf{x})\dot{x}^{i}\dot{x}^{j}=0,
\label{ELeq2}
\end{equation} 
where
\begin{equation}
\Gamma _{~~ij}^{k}=2\left( C_{ij}C^{kl}\partial _{l}\Delta
-\delta _{~~i}^{k}\partial _{j}\Delta-\delta _{~~j}^{k}\partial _{i}\Delta \right)/\Delta,
\end{equation}%
$C_{ij}\equiv (\mathbf{C})_{ij}=\mathrm{Tr}[\sigma _{i}\sigma _{j}]$,
and $C^{ij}=(\mathbf{C}^{-1})_{ij}$. (vii) Including various physical
constraints 
into the variational description of AQC is natural through the Lagrange
multiplier method.

\textit{Geometrization of AQC.---}Motivated by the existence of a
differential-geometric description for circuit optimization \cite%
{Nielsen-science}, and by the resemblance of Eq.~(\ref{ELeq2}) to a geodesic
equation \cite{Nakahara-book}, we reformulate QAB in a
differential-geometric language. This in turn endows AQC with a Riemannian
manifold structure.
An immediate advantage of this
geometrization is that it equips AQC with the powerful techniques and
tools of differential geometry, and
shows that the problem of finding the QAB can be viewed as belonging to geometric
control theory \cite{Jurdjevic}. Furthermore, geometrization allows
treating AQC and the circuit model
in a universal geometric setting \cite{Nielsen-science}, which may
suggest a natural alternative to Refs.~\cite{AQCeq} for proving the
equivalence of AQC and the circuit model.

The transition from the EL equations to a geodesic equation is
possible when one can write $\mathcal{L}'[\dot{\mathbf{x}},\mathbf{x}%
]=g_{ij}(\mathbf{x})\dot{x}^{i}\dot{x}^{j}$, 
where $\mathbf{g}$ (with matrix elements $g_{ij}$)
is a differentiable and invertible matrix \cite{Nakahara-book}. From the
definition of $\mathcal{L}$ for the QAB, we then obtain 
\begin{equation}
g_{ij}(\mathbf{x})=\mathrm{Tr}[\partial _{i}H(\mathbf{x})\partial _{j}H(%
\mathbf{x})]/\Delta ^{4}(\mathbf{x}),  \label{metric}
\end{equation}%
which is the sought-after metric tensor. In interaction
coordinates, 
for example, 
$\mathbf{g}(\mathbf{x})=\mathbf{C}/\Delta
^{4}(\mathbf{x})$, 
while it is represented differently in other coordinates \cite{Nakahara-book}.

In this framework, then, 
the QAB is equivalent to the geodesic over $(\mathcal{M},\mathbf{g})$, and Eq.~(%
\ref{ELeq2}) is the geodesic equation, in which $\Gamma _{~~ij}^{k}=\frac{1}{%
2}g^{il}(\partial_j g_{kl}+ \partial_k g_{jl}- \partial_l g_{jk})$ 
are the connection coefficients [where $g^{ij}\equiv(\mathbf{g}^{-1})_{ij}$].
Since $g_{ij}\propto \Delta^{-4}$ 
(if the numerator does not contribute a power of $\Delta$) 
we find $\Gamma \sim g^{-1}\partial g\sim \Delta
^{-1}\partial \Delta $. Using standard expressions \cite{Nakahara-book}, one
can calculate the Riemann curvature tensor $\mathbf{R}$ from the connection
coefficients and the metric tensor, yielding $\mathbf{R}\sim \partial
^{2}g+g\Gamma ^{2}\sim \Delta ^{-6}$.

\textit{Examples}.---We consider the following two-dimensional (2-d)
interpolating Hamiltonian: 
\begin{equation}
H\left( x^{1}(s),x^{2}(s)\right) =x^{1}(s)P_{\mathbf{a}}^{\perp }+x^{2}(s)P_{%
\mathbf{b}}^{\perp },  \label{LinearH}
\end{equation}%
where $P_{\mathbf{a}}^{\perp }=\openone-|\mathbf{a}\rangle \langle \mathbf{a}%
|$ for the normalized vector $|\mathbf{a}\rangle \in \mathcal{H}$ with $\dim
(\mathcal{H})=N$ (similarly for $P_{\mathbf{b}}^{\perp }$), $\alpha
_{0}\equiv \langle \mathbf{a}|\mathbf{b}\rangle $ is a known function of $N$
alone, and $x^{1}(0)=x^{2}(1)=1$; $x^{1}(1)=x^{2}(0)=0$. We can always
find $|\mathbf{a}^{\perp }\rangle $ such that $|\mathbf{b}\rangle =\alpha
_{0}|\mathbf{a}\rangle +\alpha _{1}|\mathbf{a}^{\perp }\rangle $, where $%
\langle \mathbf{a}|\mathbf{a}^{\perp }\rangle =0$, and $\alpha _{1}=\langle 
\mathbf{a}^{\perp }|\mathbf{b}\rangle $. Completing $\{|\mathbf{a}\rangle ,|%
\mathbf{a}^{\perp }\rangle \}$ to a basis for $\mathcal{H}$ we can easily
diagonalize the Hamiltonian (\ref{LinearH}), and find that the gap between
the ground state and the first excited state is $\Delta (x^{1},x^{2})=\sqrt{%
  (x^{1})^{2}+(x^{2})^{2}+2(2|\alpha _{0}|^{2}-1)x^{1}x^{2}}$.
While the general 2-d problem requires a numerical solution, we
find that if we impose a one-dimensional 
(1-d) constraint, i.e., $x(s)\equiv x^{2}(s)=1-x^{1}(s)$, then an
analytical solution is possible:
\begin{equation}
\hskip -.8mm x_{\text{QAB}}(s)=\frac{1}{2}-\frac{|\alpha _{0}|}{2\sqrt{1-|\alpha _{0}|^{2}}}%
\tan \left[ (1-2s)\arccos |\alpha _{0}|\right] .  \label{QAB-general}
\end{equation}%
We now consider two illustrative problems which are special cases of
the Hamiltonian (\ref{LinearH}).

\textit{Quantum search.---}As a first illustration we revisit
Grover's unstructured search problem, which involves finding a marked object
among $N$ objects by repeated oracle queries \cite{Nielsen-book,Grover}.
Grover's quantum circuit model solution uses 
$O(\sqrt{N})$ queries, which is
provably optimal, and a quadratic improvement over the best possible
classical strategy. This problem was successfully recast in the AQC setting
by Roland and Cerf (RC) \cite{RolandCerf}, who considered the 1-d version of
(\ref{LinearH}) with $x(s)\equiv x^{2}(s)=1-x^{1}(s)$, $|\mathbf{a}\rangle
=\sum_{k=0}^{N-1}|k\rangle /\sqrt{N}$ (equal superposition), $|\mathbf{b}%
\rangle =|m\rangle $, and the fixed index $m\in \{0,\ldots ,N-1\}$ being the
\textquotedblleft marked item\textquotedblright . Thus $\alpha _{0}=1/\sqrt{N%
}$, with $N=2^{n}$ the dimension of the Hilbert space of $n$ qubits.
It turns out that the optimal 1-d solution (\ref{QAB-general}) coincides
precisely with the solution found by RC, who proved its optimality (in the
sense of $O(\sqrt{N})$ scaling for a fixed error) without the use of
variational optimization. We now extend the analysis by considering 2-d and
4-d parametrizations, which corresponds to finding optimal curves on 2-d and
4-d manifolds, respectively. The 2-d case is given by Eq.~(\ref{LinearH})
and the discussion that follows it, with $|\mathbf{a}\rangle $ and $|\mathbf{%
b}\rangle $ as above. In the 4-d case, we first consider a general one-qubit
Hamiltonian $H(x^{1},x^{2},x^{3},x^{4})=\frac{1}{\sqrt{2}}(x^{1}\openone%
+x^{2}\sigma _{x}+x^{3}\sigma _{y}+x^{4}\sigma _{z})$, and solve the
corresponding geodesic (or QAB) differential equations. Next we recall that
Grover's search is effectively a 2-d problem 
(in the $\{|{\bf a}\rangle,|m\rangle\}$ basis).
This enables us to use the 4-d setting for
finding a Groverian geodesic path, with the proper boundary conditions
corresponding to $|{\bf a}\rangle = \sqrt{(N-1)/N} |0\rangle +
1/\sqrt{N} |1\rangle$ and, for example, $|m\rangle =|1\rangle$. 
However, note that the parametrization of $H(x^{1},x^{2},x^{3},x^{4})$ above
is not the most general 4-d parametrization when $n>1$.

The 1-d RC\ analysis employed the local adiabatic condition
(\ref{eq:T-local}) to recover the optimal scaling 
$T_{\text{opt}}\propto \sqrt{N}$, 
for $N\gg 1$. This might suggest that there is no room for further improvement, but we
recall that in the AQC\ setting the fidelity is $1$ only in the limit $%
T\rightarrow \infty $. Thus we compare the error $\delta (T)$ for the RC
interpolation to the error obtained from the optimal 2-d interpolation. The
result for
$n=6$ qubits is shown in Fig.~\ref{comparison} (left); results
for other values of $n$ are qualitatively similar, though the advantage of
the optimal interpolation shrinks as $n$ grows. 
\emph{The optimal 2-d interpolation results in a smaller error for most
values of} $T$, \emph{a tendency that increases as} $T$ \emph{grows}.
Conversely, for
most values of the
error $\delta $ the 2-d QAB requires a smaller time $%
T$ than the RC curve. 
The middle panel shows the further improvement resulting from the 4-d
interpolation. These
results provide a rather striking demonstration of the power of our
formalism, as due to its highly optimized nature, the Grover example is one
where hardly any improvement was to be expected.

\begin{figure}[tp]
\includegraphics[width=4cm,height=2.95cm]{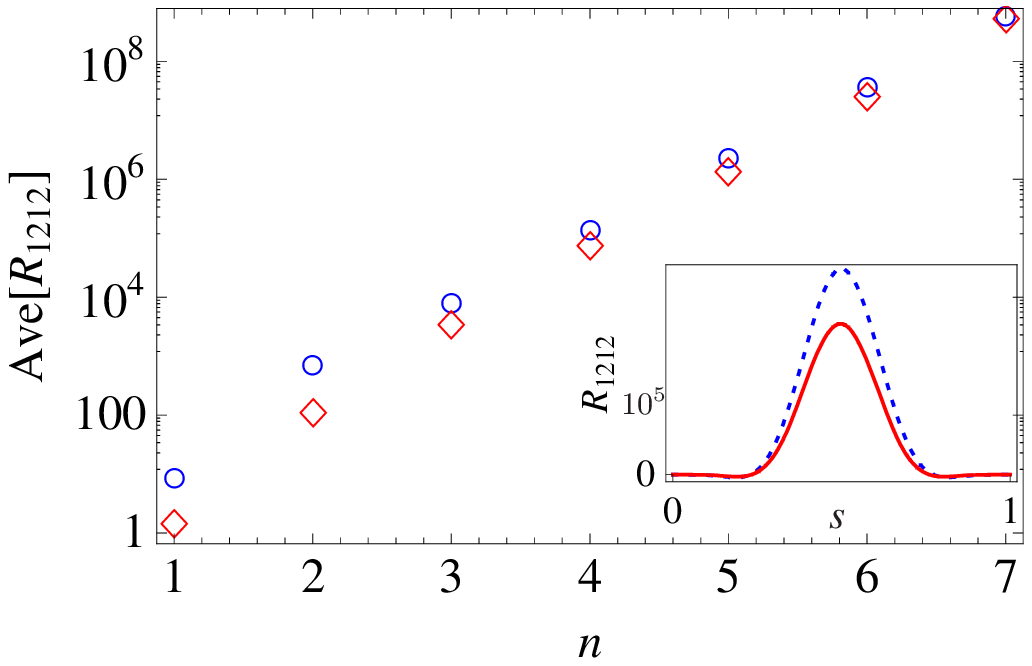}\hskip2mm
\includegraphics[width=4.3cm,height=2.95cm]{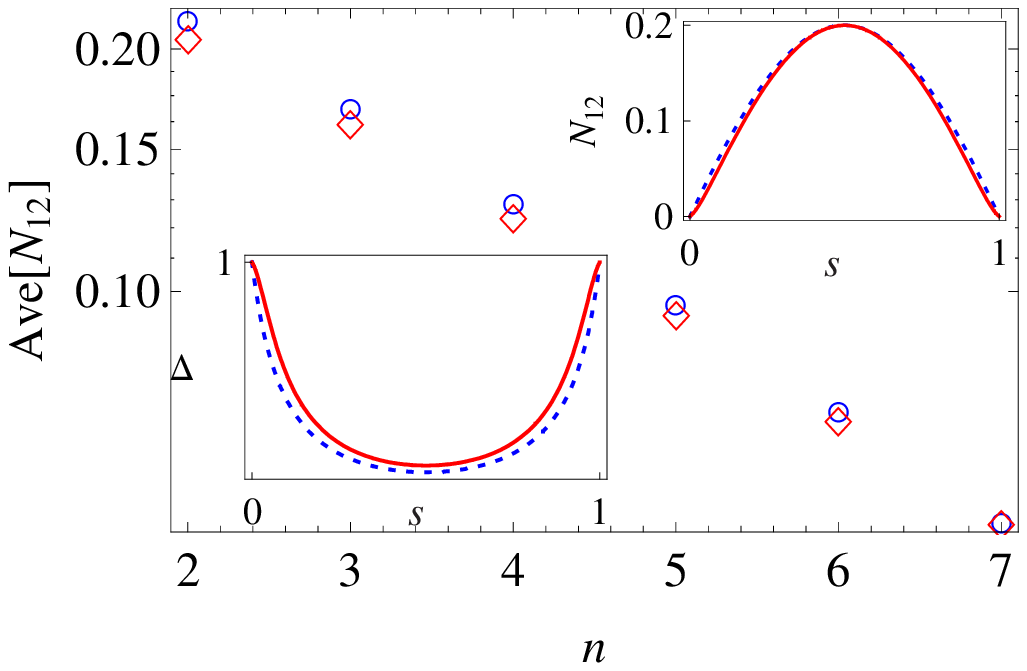}
\caption{(color online) Red diamonds or solid line  
(blue circles or dashed line) represent the 2-d QAB (RC)
path. Left: Average curvature component $R_{1212}$ vs
$n$. Inset: instantaneous curvature for $n=4$. Results for other values of $n$ are
qualitatively similar. Right: Same for negativity $N_{12}$
\protect\cite{neg} -- between qubits $1$ and $2$. 
Inset: instantaneous gap (lower left) and negativity (upper right) for $n=4$. 
Here, $\text{Ave}_{\gamma}[X] \equiv \int_{\gamma} X[\mathbf{x}(s)] \mathrm{d}s/\int_{\gamma}
\mathrm{d}s$ for given path $\gamma$; $X\in\{N_{12},R_{1212}\}$ and
$\gamma \in \{{\rm RC,QAB}\}$.
Note that same values
of $s$ (\textit{local} time)
do not necessarily indicate the same real (\textit{proper})
time $t$ for different paths.
}
\label{NR}
\end{figure}

Figure~\ref{comparison} (right) depicts the RC 
and 2-d optimal curves over the curvature $R_{1212}$ surface. 
Clearly, the optimal curve follows a path of lower curvature. 
This is confirmed in Fig.~\ref{NR} (left), for different values of $n$. Figure~\ref{NR} 
(right) depicts the amount of bipartite entanglement
along the RC and 2-d optimal paths. {\em In spite of its improved performance, 
there is less entanglement along the 2-d optimal path than along the
RC path},
so that more entanglement does not always translate into higher algorithmic efficiency. We have
verified (not shown) that the same picture emerges in terms of the entanglement
entropy (or block entanglement) \cite{Vidal}. The explanation for this lower entanglement along the 
QAB 
is that it has a larger instantaneous gap than the RC path [Fig.~\ref{NR} (right)].
Indeed, it has been shown that for Grover's algorithm the entanglement entropy 
is small away from the finite-size precursor of the first-order QPT, but peaks near the
critical point \cite{AQCmanybody}, and we have verified the same for
the 2-d QAB. 
Finally, the reason that QAB follows a path with larger gap is
that this is consistent with higher adiabaticity. By our
previous scaling result $\mathbf{R}\sim \Delta ^{-6}$, it is also
consistent with lower curvature.

\textit{Linear equations.---}Solving linear equations of the type $A\mathbf{y%
}=\mathbf{a}$, 
where $A$ is a given (Hermitian) $N\times N$ matrix and $%
\mathbf{a}$ is a given
vector, is a common problem. Recently
a quantum algorithm was proposed in the circuit model that can obtain $%
\mathbf{y}$ for well-conditioned, \emph{sparse} matrices in a time scaling
as $\text{polylog}(N)$ \cite{Harrow}. Here we consider the problem of
finding $\mathbf{y}=A^{-1}\mathbf{a}$ as one of oracular \textit{adiabatic
state generation} \cite{ASG}. To do so we let $|\mathbf{b}\rangle =A^{-1}|%
\mathbf{a}\rangle /\Vert A^{-1}|\mathbf{a}\rangle \Vert $. The formulation
given above for the Hamiltonian (\ref{LinearH}) then applies. For
concreteness we let $|\mathbf{a}\rangle =(1,\ldots ,1)^{T}/\sqrt{N}$ and
take $A$ to be an $N\times N$ Toeplitz matrix
whose first row and column are successive natural numbers, starting
from 1. Toeplitz matrices
have important applications in signal processing \cite{Toeplitz}, and
are not sparse. We then find $\alpha _{0}=\sqrt{2/N}$ and hence can deduce
immediately -- by analogy to the Grover case, where $\alpha _{0}=\sqrt{1/N}$ --
that the optimal 1-d interpolation will give rise to a run-time $%
T$ scaling as $O(\sqrt{N})$ for a fixed error.
Moreover, 2-d and 4-d interpolations will further improve the
error at fixed run-time.
It is interesting to note that the most
efficient known classical algorithm for inverting an $N\times N$ Toeplitz
matrix requires $O(N\log ^{2}N)$ steps \cite{Toeplitz}, though a direct
comparison is not possible due to our oracular setting.

\textit{Conclusion and outlook.---}We have presented a time-optimal,
differential-geometric framework for AQC, and
discussed its implications for the optimal design of adiabatic algorithms.
The power of this new framework was illustrated via an example showing
how the performance
of an adiabatic algorithm can be improved by increasing the dimension of the
control parameter space, and how geometrization sheds light on the role
of entanglement and control manifold curvature in
this enhanced performance.
The method presented here is general
and can in principle be used to optimize any adiabatic quantum
algorithm for which the gap (or estimate thereof) is known.
An important next step is to
incorporate decoherence-mitigation strategies \cite{Lidar-FTAQC}.

\textit{Acknowledgments}.---Supported by the NSF under grants No.
CCF-726439, PHY-802678, PHY-803304, and PHY-0803371, and by NSERC and MRI.


\end{document}